\documentclass[%
 reprint,
superscriptaddress,
preprintnumbers,
nofootinbib,
twocolumn,
 amsmath,amssymb,
 aps,prl,
floatfix,
]{revtex4-2}
\usepackage{multirow}
\usepackage{graphicx}
\usepackage{dcolumn}
\usepackage{bm}
\usepackage{hyperref}
\usepackage{bbm}
\usepackage{wasysym}
\usepackage{xcolor,soul}

\usepackage{xargs}
\usepackage[colorinlistoftodos,prependcaption,textsize=footnotesize]{todonotes}
\newcommandx{\DM}[2][1=]{\todo[linecolor=orange,backgroundcolor=orange!25,bordercolor=orange,#1]{DM: #2}}
\newcommandx{\KH}[2][1=]{\todo[linecolor=green,backgroundcolor=green!25,bordercolor=green,#1]{KH: #2}}
\newcommandx{\UW}[2][1=]{\todo[linecolor=blue,backgroundcolor=blue!25,bordercolor=blue,#1]{UW: #2}}
\newcommandx{\AI}[2][1=]{\todo[linecolor=red,backgroundcolor=red!25,bordercolor=red,#1]{AI: #2}}

\usepackage[export]{adjustbox}

\usepackage{changes}

\begin{document}

\title{Machine-learned RG-improved gauge actions and classically perfect gradient flows}

\author{Kieran Holland}
\email{kholland@pacific.edu}
\affiliation{University of the Pacific,
3601 Pacific Avenue, Stockton, California 95211, USA}

\author{Andreas Ipp}
\email{ipp@hep.itp.tuwien.ac.at}
\affiliation{Institute for Theoretical Physics, TU Wien, Wiedner Hauptstraße  8-10/136, A-1040 Vienna, Austria}

\author{David I.~M\"{u}ller}
\email{dmueller@hep.itp.tuwien.ac.at}
\affiliation{Institute for Theoretical Physics, TU Wien, Wiedner Hauptstraße  8-10/136, A-1040 Vienna, Austria}

\author{Urs Wenger}
\email{wenger@itp.unibe.ch}
\affiliation{Albert Einstein Center for Fundamental Physics,
Institute for Theoretical Physics, University of Bern, Sidlerstraße 5, CH-3012 Bern, Switzerland}

\begin{abstract}
Extracting continuum properties of quantum field theories from discretized spacetime is challenging due to lattice artifacts. Renormalization-group (RG)-improved lattice actions can preserve continuum properties, but 
are in general difficult to parameterize. Machine learning (ML) with gauge-equivariant convolutional neural networks provides a way to efficiently  describe such actions. We test a machine-learned RG-improved lattice gauge action, the classically perfect fixed-point (FP) action, for four-dimensional SU(3) gauge theory through Monte Carlo simulations. 
We establish that the gradient flow of the FP action is free of tree-level discretization effects to all orders in the lattice spacing, making it classically perfect. This allows us to test the quality of improvement of the FP action, without introducing additional artifacts. We find that discretization effects in gradient-flow observables are highly suppressed and less than 1\% up to lattice spacings of 0.14 fm, allowing continuum physics to be extracted from coarse lattices. The quality of improvement achieved motivates the use of the FP action in future gauge theory studies.  The advantages of ML-based parameterizations also highlight the possibility of realizing quantum perfect actions in lattice gauge theory.
\end{abstract}

\maketitle

\date{\today}

\textit{Introduction} ---
The Standard Model of particle physics gives excellent agreement between experimental measurements and theoretical predictions to very high precision. A central element is Quantum Chromodynamics (QCD), the quantum field theory of quark-gluon interactions, where Monte Carlo (MC)  simulations play a crucial role in testing nonperturbative dynamics~\cite{FlavourLatticeAveragingGroupFLAG:2021npn}. A necessary step in lattice simulations is spacetime discretization, which ultimately must be removed to obtain physical continuum predictions. However, the extrapolation to the continuum limit is often the dominant source of uncertainty in lattice studies and remains a serious challenge: reducing the lattice spacing $a \rightarrow 0$, simulations become less efficient~\cite{Schaefer:2010hu}, requiring longer Markov chains for high statistical accuracy (\textit{critical slowing down}) and to fully explore the phase space (\textit{topological freezing}). These systematic effects are a barrier to increasing the precision of lattice predictions for the Standard Model. To overcome this hurdle, one can use improved lattice actions~\cite{Symanzik:1983dc,Symanzik:1983gh,Luscher:1984xn}, designed to remove the leading lattice artifacts, often through a perturbative expansion in $a$. The efficacy of the improvement must be empirically tested via MC simulations. 
The problem has also stimulated the study of various alternatives, such as normalizing flows, which use machine learning (ML) to transform random samples to target distributions~\cite{Kanwar:2020xzo,Abbott:2022zhs,Abbott:2024kfc,DelDebbio:2021qwf,Albandea:2023wgd,Bacchio:2022vje,Gerdes:2022eve,Gerdes:2024rjk,Nicoli:2023qsl}. 
However, normalizing flows face significant challenges, such as achieving high efficiency in four-dimensional quantum field theories—measured by effective sample size—and ensuring that models trained at one lattice spacing can generalize effectively to generate ensembles at other lattice spacings. 
Combined with the persistent difficulties of extrapolating to the continuum limit, these challenges underscore the urgent need for new methods that can achieve accurate results with small lattice artifacts already on coarse lattices, thus evading critical slowing down and topological freezing.

In this Letter, we employ an alternative approach, based on fixed-point (FP) lattice actions \cite{Hasenfratz:1993sp} which are implicitly and nonperturbatively defined using the properties of Wilson's renormalization group (RG) for asymptotically free theories \cite{Wilson:1973jj,Wilson:1974mb,Wilson:1983xri}. It is used to extract continuum quantities from coarse lattices with significantly reduced lattice artifacts. We demonstrate the effectiveness of the FP action in MC simulations of four-dimensional SU(3) gauge-field theory, parameterized using a convolutional neural network with exact gauge invariance \cite{Favoni:2020reg}.  This result is made possible by our new theoretical insight, presented here, that FP actions yield a classically perfect gradient flow, i.e., without tree-level lattice artifacts to all orders in the lattice spacing. This result also builds on the success of our recent work~\cite{Wenger:2025sre,Holland:2024muu,Holland:2023ews}, where we demonstrated that ML-based parameterizations can accurately capture the FP action across a wide range of gauge-field fluctuations---from very fine to very coarse lattice spacings. The resulting parameterization achieves an average relative error in the action value of less than $0.2\%$, representing an improvement by over an order of magnitude compared to previous approaches developed earlier \cite{DeGrand:1995ji,Blatter:1996np,Niedermayer:2000yx}. The ML framework also enables accurate parameterization of the derivatives of the FP action w.r.t.~the gauge fields, which are crucial for both efficient network training and the practical implementations of MC algorithms. It is the first successful application of a machine-learning technique to the simulation of a four-dimensional gauge-field theory which scales to sufficiently large lattice volumes and small lattice spacings to allow for controlled continuum limits of physically relevant quantities. As such it is of general interest both for the broad physics community as well as for the machine-learning community. The machine-learned FP action can be generally used in any lattice-QCD study and provides a new way to reduce discretization artifacts on coarse lattices, making it broadly useful. 

Apart from the machine-learning aspect, the classically perfect properties of the FP gradient flow (GF) are of general interest, because it finds applications in many physical fields, e.g., statistical mechanics, fluid dynamics, diffusion processes, optimization problems, and image reconstruction.

{\it Renormalization group and fixed-point actions ---}
For SU($N$) gauge theories, an RG transformation of a lattice action $\mathcal A$ from a fine to a coarse lattice can be defined through  
\begin{equation}
    \exp(-\beta' \mathcal A'[V]) = \int \mathcal{D} U \exp( -\beta \{ \mathcal A[U] + \mathcal{T}[U,V] \} ),
    \label{eq:rgt}
\end{equation}
where $\beta=2N/g^2$ and the fine lattice configuration $U$ is connected to the coarse configuration $V$ through the RG blocking kernel
\begin{equation}
    \mathcal{T}[U,V] = - \kappa \sum_{n_B,\mu} \Bigl\{ {\rm Re Tr} (V_{n_B,\mu}Q_{n_B,\mu}^\dagger[U] ) - \mathcal{N}^\infty_\mu \Bigr\}.
    \label{eq:block}
\end{equation}
The variables $Q_{n_B,\mu}$ are blocked links constructed from the fine links $U_{n,\mu}$ and the normalization constant $\mathcal{N}^\infty_\mu = \max_{W \in \text{SU}(N)} \left[ {\rm Re Tr} (W Q_{n_B,\mu}^\dagger) \right]$ guarantees the invariance of the partition function after an RG blocking step. The blocking details and parameter $\kappa$ can be freely chosen to maximize the compactness of the FP action \cite{Blatter:1996np} and in this work we follow the RG blocking therein.

In the limit $1/g^2 \rightarrow \infty$ of asymptotically free quantum field theories, the FP of the RG is scale-invariant and defines the continuum limit of the gauge theory. Moreover, the path integral on the RHS of Eq.~(\ref{eq:rgt}) reduces to a saddle point problem yielding the classical FP equation \cite{Hasenfratz:1993sp}
\begin{equation}
    \mathcal A^{\rm FP}[V] = \min_{ \{U\} } \left[ \mathcal A^{\rm FP}[U] + \mathcal{T}[U,V] \right].
    \label{eq:fp}
\end{equation}
The resulting FP action $\mathcal{A}^{\rm FP}$ is {\it classically perfect}, i.e., it has no lattice artifacts at tree level and exactly reproduces continuum classical properties at finite
lattice spacing, e.g., dispersion relation~\cite{Hasenfratz:1993sp}, topology~\cite{DeGrand:1995ji,Blatter:1995ik}, or index theorem~\cite{Hasenfratz:1998ri}, chirality~\cite{Hasenfratz:1998jp,Niedermayer:1998bi} when fermions are included. A key theoretical result of this work is the derivation that the gradient flow of the FP action is also classically perfect.

\textit{Classically perfect gradient flow} ---
To rigorously test the quality of improvement in any lattice action, it is essential to use precisely measurable observables that avoid introducing significant cutoff effects. Observables based on the gradient flow (GF) of gauge fields~\cite{Luscher:2010iy} are a natural choice, given their small systematics and high statistical accuracy. In addition, due to the  renormalization properties of the GF~\cite{Luscher:2011bx}, these observables are widely used for the setting of physical scales and the measurement of the $\beta$-function of the gauge coupling. We show here that the GF based on the FP action has the additional desirable property of being {\em classically perfect}: it has no tree-level discretization effects to any order in the lattice spacing $a$. 

In continuum language, for an action ${\cal A}$ the GF evolves the non-Abelian gauge field $A_\mu(t)$ along a fictitious flow time $t$ according to\added{~\cite{Luscher:2010iy, Luscher:2011bx,Narayanan:2006rf,Lohmayer:2011si}}
\begin{equation}
    \frac{d A_\mu(t)}{dt} = - \frac{\delta {\cal A}}{\delta A_\mu(t)},
\end{equation}
where the starting value $A_\mu(t=0)$ is the original gauge field and the action is ${\cal A} = (1/g^2) \int d^4x (- \frac{1}{2} {\rm Tr} F_{\mu \nu}F_{\mu \nu})$, with field strength $F_{\mu \nu} = \partial_\mu A_\nu - \partial_\nu A_\mu + [A_\mu,A_\nu]$ and gauge coupling $g$.
The action density $E(t)=- \frac{1}{2} {\rm Tr} F_{\mu \nu}F_{\mu \nu}$ at flow time $t$ is a renormalized quantity, and for SU($N$) gauge theory it can be connected perturbatively to a renormalized coupling $g$ in a given scheme, e.g., the $\overline{\rm MS}$ scheme, as\added{~\cite{Luscher:2011bx}}
\begin{equation}
    t^2 \langle E(t) \rangle = \frac{3(N^2-1)g^2}{128 \pi^2}\left( 1 + {\cal O}(g^2) \right),
\end{equation}
where $\langle\cdot\rangle$ denotes the gauge-field average. 

On the lattice, the perturbative expansion is in the bare coupling $g_0$ with\added{~\cite{Fodor:2014cpa}} 
\begin{equation}
t^2 \langle E(t) \rangle = \frac{3(N^2-1)g_0^2}{128 \pi^2}\left( C(a^2/t) + {\cal O}(g_0^2) \right), 
\end{equation}
where
\begin{multline}
        C(a^2/t) 
        = \frac{64 \pi^2 t^2}{3} \times\\
        \int_{-\pi/a}^{\pi/a} \frac{d^4 p}{(2 \pi)^4} {\rm Tr}\left[e^{-t({\cal S}^f+{\cal G})}({\cal S}^g+{\cal G})^{-1}e^{-t({\cal S}^f+{\cal G})} {\cal S}^e\right]
\end{multline}
contains the tree-level lattice artifacts as deviations from 1. Here, the action density ${\cal S}$ is in momentum space to quadratic order in the fields, and ${\cal S}^e, {\cal S}^f, {\cal S}^g$ correspond to the (possibly different) choices of the discretized action for the action-density observable $E$, for the GF, and for the MC simulation, respectively, while ${\cal G}$ is a gauge fixing term. For any choice of action discretizations, the value of $C(a^2/t)$ can be calculated and used to define a tree-level improved coupling at finite lattice spacing~\cite{Fodor:2014cpa} (see also~\cite{Ramos:2015baa}). Using the same lattice action for all three gives
\begin{equation}
    C(a^2/t) = \frac{64 \pi^2 t^2}{3} \int_{-\pi/a}^{\pi/a} \frac{d^4 p}{(2 \pi)^4} {\rm Tr}\left[e^{-2t({\cal S}+{\cal G})}\right],
\end{equation}
since the various factors under the trace commute with each other.

For the exact FP action ${\cal A^{\rm FP}}$, the gluon propagator has poles $1/(p+2\pi l)^2$ for momenta $-\pi/a \le p_\mu \le \pi/a$ and all possible integers $l$~\cite{DeGrand:1995ji}, since the iterative RG connection to finer and finer lattices extends the momentum integration range to the full interval 
$\pm \infty$ even at finite lattice spacing, producing precisely the continuum dispersion relation. Hence, for ${\cal S}^{\rm FP}$ one obtains
\begin{equation}
    C(a^2/t) = \frac{64 \pi^2 t^2}{3} \cdot 3 \left(\int_{-\infty}^{+\infty} \frac{dp}{2 \pi} e^{-2t p^2} \right)^4 
    = \frac{64 \pi^2 t^2}{(\sqrt{8 \pi t})^4} = 1, \nonumber
\end{equation}
i.e., the exact FP action has no tree-level cutoff dependence in $t^2 \langle E(t) \rangle$; lattice artifacts appear in loop corrections, i.e., as quantum effects, at ${\cal O}(a^2 g_0^2)$. This is a new example of FP actions keeping continuum classical properties intact at finite lattice spacing.

\textit{Machine learning architecture} ---
The practical limitations in using the FP action in MC simulations arise from two factors: the size of quantum lattice artifacts, and the inherent imperfections in the fixed-point parameterization for four-dimensional SU(3) gauge theory. The latter is addressed here using a gauge-equivariant convolutional neural network \cite{Favoni:2020reg} and ML techniques.
The ML architecture designed for the FP action is described in detail in~\cite{Holland:2024muu}. In brief, the main building block of the  neural network is a bilinear convolution which computes various products of Wilson loops, starting from $1 \times 1$ plaquettes and the links $U_{n,\mu}$ of a gauge-field configuration as input. Gauge covariance is exactly maintained throughout by accounting for parallel transport in the convolution.  As the network depth increases, larger and more complex closed loops are generated. The final layer is a trace giving the action value as output. The network respects translational symmetry and is (ultra-)local, which allows it to be used on any lattice size. Additionally, our architecture is set up to reproduce the Yang-Mills action in the naive continuum limit and approximates a perturbatively defined FP action on smooth fields \cite{Blatter:1996np}, see \cite{Holland:2024muu} for further details. 

The goal of training a model is to approximate the action values according to the FP equation in Eq.(\ref{eq:fp}). A crucial element is that the exact derivatives of ${\cal A}^{\rm FP}$ with respect to gauge links $U_{n,\mu}$ can be easily obtained from the FP equation, with each gauge configuration giving $8 \times 4 \times$(volume) data for learning. In the network, the derivatives can be calculated exactly via backpropagation. The loss function for training hence contains both action values and derivatives. The best architecture has been obtained from a hyperparameter scan and consists of three gauge-covariant convolutional layers. It has been trained on equilibrated configurations for a variety of lattice sizes and a large range of coupling parameters from very smooth to coarse gauge-field fluctuations, and has further been improved by finetuning on instanton configurations. Our best model reaches an average relative error in the action value of less than 0.2\% with the exact derivatives more accurately reproduced than in previous parameterizations \cite{Niedermayer:2000yx} across all lattice spacings. The fidelity of the model for very smooth configurations is particularly important for the GF because the gauge-field fluctuations become very smooth at large flow time.

\textit{Fixed-point action simulations} ---
The availability of accurate gauge-field derivatives from our ML parameterization model makes the Hybrid Monte Carlo (HMC) algorithm a suitable choice for simulations, where artificial conjugate momenta are introduced to update the gauge links through molecular dynamics~\cite{Duane:1987de,Gottlieb:1987mq}. The simulations can be accelerated with a coarser choice of time step when solving the equations of motion~\cite{Omelyan:2003bjg} using in particular the 4MN4FP variant of Omelyan integration scheme~\cite{Takaishi:2005tz}. The GF for gauge links $dU_\mu/dt = - i (\delta {\cal A}^{\rm FP}/\delta U_\mu) U_\mu$ is solved using the third-order Runge-Kutta integrator~\cite{Luscher:2010iy}. The ML network is memory intensive, but the memory usage only scales linearly with the lattice volume; the largest lattice volumes we simulate are $18^4$, for which we can generate a new HMC trajectory in $\sim 4$ minutes on one NVidia A100 SXM6 64GB GPU. With the FP action, a smaller lattice volume is compensated by coarser lattice spacing to yield near-continuum-physics results in large enough physical volume. As a consistency check of continuum predictions with the FP action, we also simulate four-dimensional SU(3) gauge theory with the publicly available {\tt openqcd} package~\cite{openqcd} using either Wilson or tree-level Symanzik improved gauge actions and measuring with Wilson GF.\footnote{
This generic choice, instead of the alternative schemes in Refs.~\cite{Fodor:2014cpa,Ramos:2015baa}, is justified by the fact that we are first of all interested in the continuum values which are universal and independent of the choice. Moreover, when comparing lattice artifacts, the improvement
schemes 
are highly problem-specific, while the FP approach is general and applies to other observables.}

\textit{Scaling tests} ---
The GF provides several physically relevant quantities that can be extracted. For example, a reference scale $t_c/a^2$ can be set through the condition $t_c^2 \langle E(t_c) \rangle = c$, where a common choice is $c=0.3$~\cite{Luscher:2010iy}. An alternative scale $w_d/a$ is set through the derivative condition
\begin{equation}
    t \cdot \frac{d}{dt} \left[ t^2 \langle E(t) \rangle\right]_{t=t_d} = d, \hspace{5mm} w_d = \sqrt{t_d},
\end{equation}
where $d=0.3$ is also frequently used~\cite{BMW:2012hcm}. 
These scales essentially probe the gauge-field fluctuations at specific
distances parameterized by the parameters $c$ and $d$.
At finite lattice spacing, the values of $t_c/a^2$ and $w_d/a$ are determined by the choice of bare coupling $6/g_0^2$ at which MC simulations are performed, as well as the action discretization used in the GF and the action density. Ratios of scales, e.g., $t_c/w_d^2$ or $t_{d}/t_c$, are physical properties, which have a well-defined value in the continuum limit independent of the discretizations employed. The necessary consistency in the continuum limit is a consequence of universality and provides a stringent test of the FP simulation results.

\begin{figure}[h!]
    \centering
    \includegraphics[width=1.\linewidth]{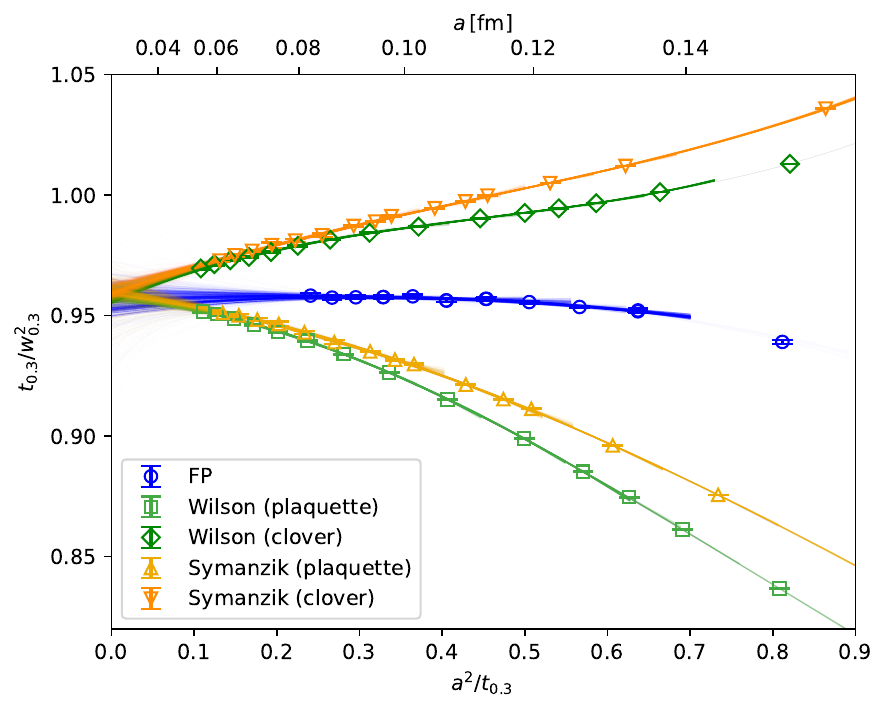}
    \includegraphics[width=1.\linewidth]{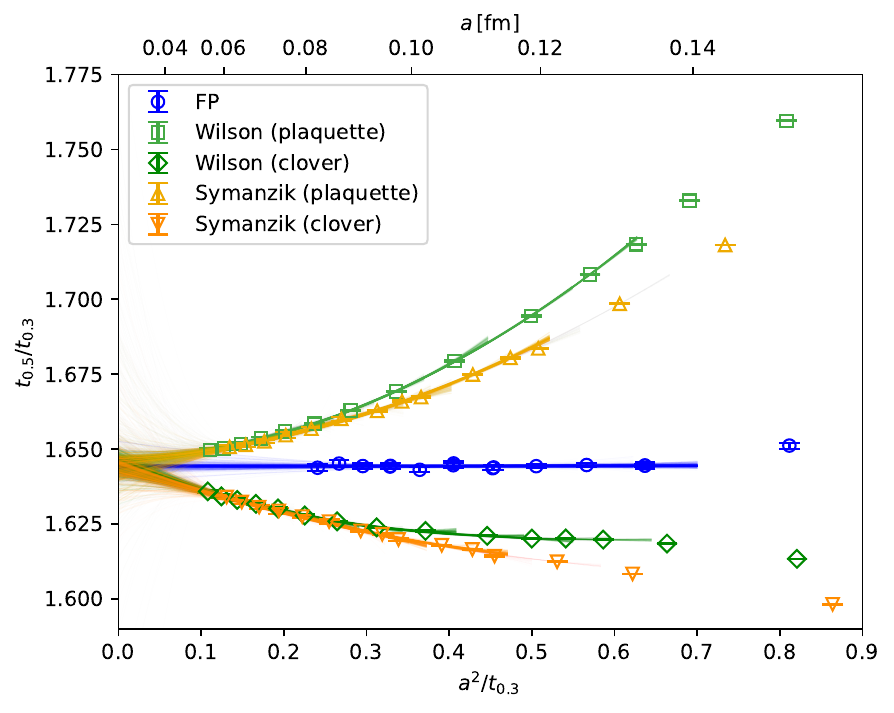}
    \caption{Continuum-limit extrapolations for the ratios $t_{0.3}/w_{0.3}^2$ and $t_{0.5}/t_{0.3}$. Results from Wilson and Symanzik MC simulations are shown using plaquette and clover discretizations of the action density. 
    } 
    \label{fig:t0w0ratio}
\end{figure}

In addition, the renormalizability of the GF action density $E(t)$ allows one to define a scheme for a renormalized coupling. For SU(3) gauge theory the GF scheme is defined by
\begin{equation}
    t^2 \langle E(t) \rangle \doteq \frac{3 g_{\rm GF}^2(t)}{16 \pi^2},
\end{equation}
where the GF renormalized coupling $g^2_{\rm GF}(t)$ runs with the RG scale $\mu = 1/\sqrt{8t}$. The $\beta$-function of the gauge coupling $\mu^2 \cdot dg^2_{\rm GF}/d(\mu^2) = -t \cdot dg^2_{\rm GF}/dt$ can be measured directly from the GF at the desired value of the coupling~\cite{Fodor:2017die,Hasenfratz:2019hpg}. For example, the choice $c=0.3$ corresponds to strong coupling $g^2_{\rm GF} \approx 15.79$. In the perturbative regime, it can be compared with the known 3-loop calculation~\cite{Harlander:2016vzb}. 

\begin{figure}
    \centering
    \includegraphics[width=1.\linewidth]{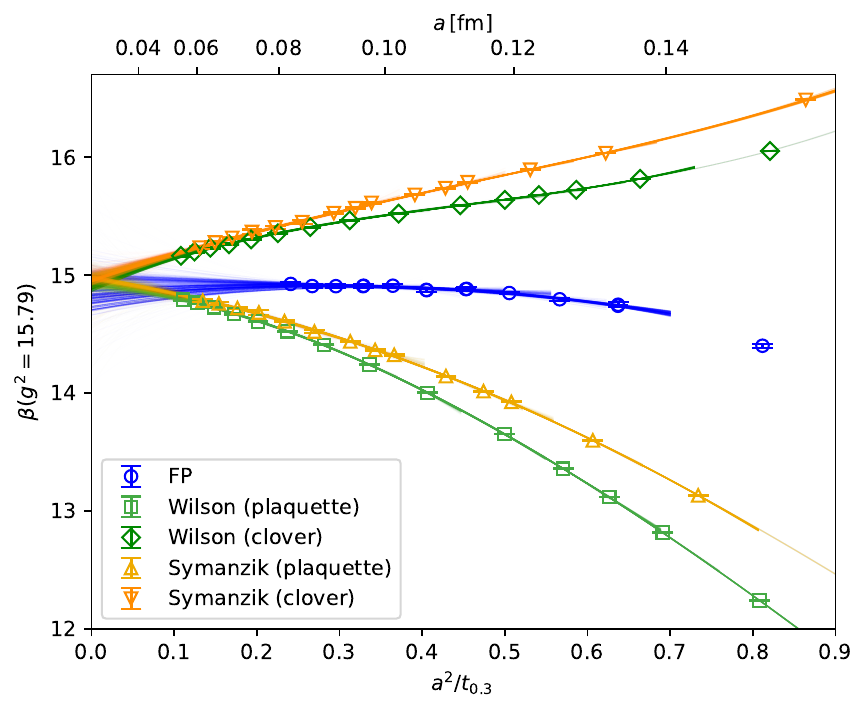}
    \caption{The $\beta$-function of the GF renormalized coupling at $g_{\rm GF}^2 = 15.79$ with highly suppressed lattice artifacts in the results for the FP compared to the Wilson and Symanzik gauge actions. }
    \label{fig:betafn}
\end{figure}

Repeating the procedure for a set of gauge ensembles at different bare couplings, the continuum limit is taken as $a^2/t_c \rightarrow 0.$ We present in Figures~\ref{fig:t0w0ratio} and~\ref{fig:betafn} results for the ratios $t_{0.3}/w_{0.3}^2$ and $t_{0.5}/t_{0.3}$, and the $\beta$-function $t \cdot dg_{\rm GF}^2/dt$ at fixed $g^2_{\rm GF}\approx 15.79$, using FP, Wilson and Symanzik action simulations. Further examples can be found in the Supplemental Material. The leading lattice artifacts are expected to be ${\cal O}(a^2)$ for the Wilson data, ${\cal O}(a^4)$ for the Symanzik data, and ${\cal O}(g^2 a^2)$ for the FP data, and we show various possible fits (polynomial in $a^2/t_{0.3}$) describing the continuum-limit extrapolations.\footnote{We note that the leading ${\cal O}(a^4)$ behaviour of the Symanzik data is masked by the ${\cal O}(a^2)$ discretization effects from the flow action (Wilson) and the action density (plaquette and clover).} The fit curves are shaded according to their Akaike Information Criterion (AIC) weight~\cite{Akaike:1974vps}, cf.~Supplemental Material.

The FP data for $t_{0.3}/w_{0.3}^2$ and the $\beta$-function show mild variation with lattice spacing, with artifacts smaller than $1\%$ at $a \simeq 0.14$ fm, using $\sqrt{t_{0.3}} = 0.1679\, \mathrm{fm}$~\cite{Giusti:2018cmp} to convert from lattice to physical scales. The ratio $t_{0.5}/t_{0.3}$ has essentially no lattice-spacing dependence below $a \lesssim 0.14$~fm using the FP action. By comparison, when discretization errors are larger, finer lattice spacings and larger lattice volumes are necessary to reach a scaling window of ${\cal O}(a^2)$ behavior while simultaneously keeping finite-volume effects small.\footnote{We note that a more meaningful discussion of the relative cost of simulations using  standard and FP actions is a delicate and difficult task as it depends heavily on the chosen benchmarks and targets for comparison, e.g., observables, level of lattice artifacts, architecture-specific code optimization, etc., and is hence not attempted here.}

To estimate systematic effects in the continuum extrapolation, such as choosing the lattice-spacing range over which to fit or the order of the fit polynomial in $a^2/t_{0.3}$, we use a procedure based on the AIC, cf.~Supplemental Material.

\begin{figure}
    \centering
    \includegraphics[width=1\linewidth]{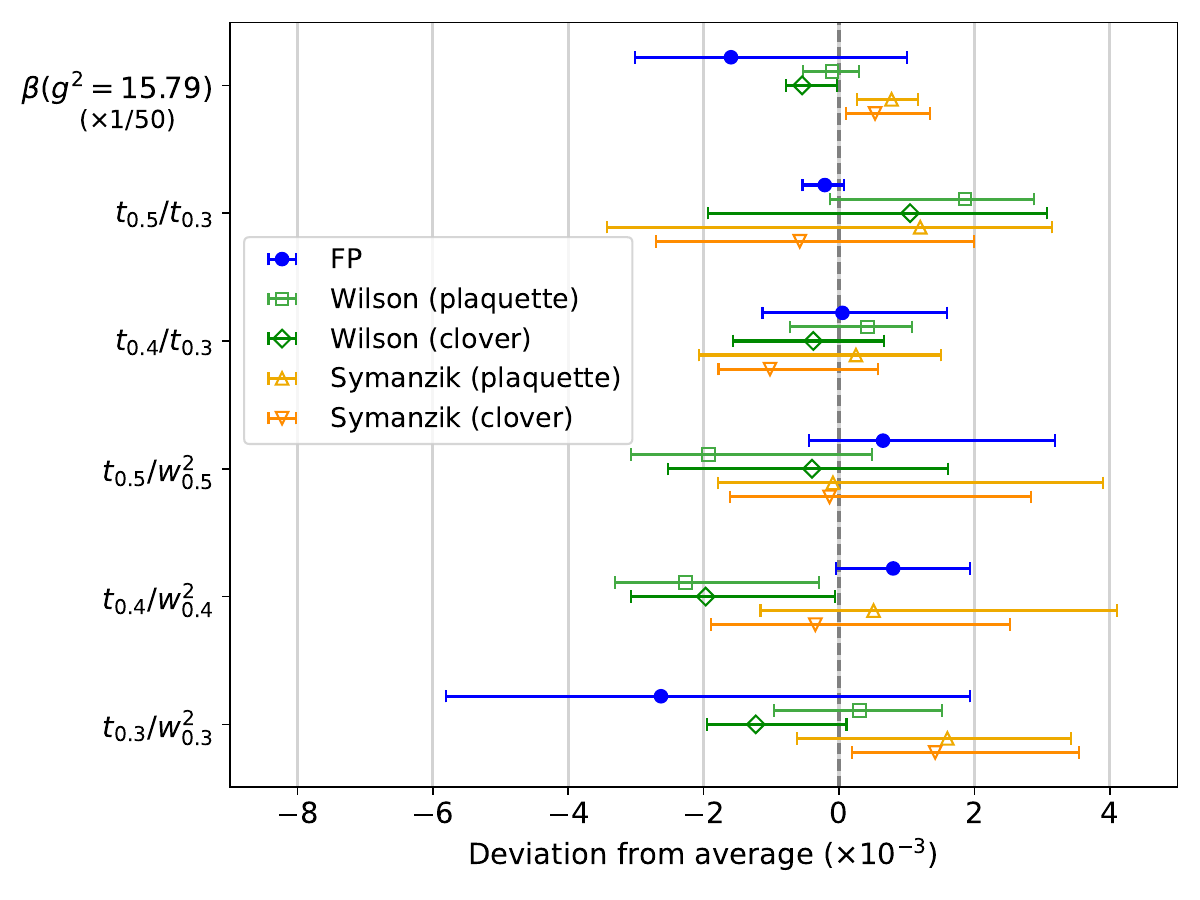}
    \caption{The comparison of continuum predictions for four-dimensional SU(3) gauge theory from MC simulations using either the FP, Wilson or tree-level Symanzik improved lattice action shows very good consistency. The $\beta$-function results are rescaled by a factor of 50 for visibility. 
    }
    \label{fig:FPwilson}
\end{figure}

The advantageous features of the FP action are visible in other choices of $c$ and $d$, cf.~Supplemental Material for further examples. In each case, FP lattice artifacts are small, and in absolute value the FP continuum predictions are accurate to per mille level. We summarize a comparison of our continuum results using the FP, Wilson or Symanzik action in Figure~\ref{fig:FPwilson}, for values of $c$ and $d$ employed by other groups~\cite{Asakawa:2015vta,Sommer:2014mea,Knechtli:2017xgy,RamosMartinez:2023tvx,Wong:2023jvr,Hasenfratz:2023bok,Brambilla:2023fsi,Spiegel:2024dec}.  There is clear consistency between FP, Wilson and Symanzik results. Simulating on smaller coarser lattices, one can measure precisely the continuum properties of the gauge theory with the FP approach.

\textit{Conclusions and outlook} ---
The explicit form of the exact RG-improved FP action is unknown and must be parameterized as a solution of the nonperturbative FP equation. Our previous study~\cite{Holland:2024muu} demonstrated that ML networks can describe the FP action for four-dimensional SU(3) gauge theory to much higher accuracy than previously possible~\cite{DeGrand:1995jk,Blatter:1996np,Niedermayer:2000yx}. In this work, we show that the FP action defines a {\it classically perfect gradient flow}, i.e., without tree-level lattice artifacts to all orders in the lattice spacing. As a consequence, the FP action successfully reduces lattice artifacts in gradient-flow observables, paving the way for various future applications. As an RG-improved lattice action, it can be used in any lattice-QCD study. One application is a new measurement of the $\Lambda$-parameter of SU(3) pure gauge theory, as a consistency check of recent results~\cite{DallaBrida:2019wur,Hasenfratz:2023bok,Wong:2023jvr} which differ from the older literature average~\cite{FlavourLatticeAveragingGroupFLAG:2021npn}, and as quantitative input for the determination of the strong coupling $\alpha_S(M_Z)$ through the decoupling of quarks~\cite{DallaBrida:2022eua}. Another possibility is using ML to  parameterize the FP Dirac operator~\cite{Hasenfratz:1998jp,Hasenfratz:2000xz,Hasenfratz:2002rp} and reduce cutoff effects in lattice-QCD simulations with dynamical quarks. A third is using ML to construct quantum perfect actions~\cite{Hasenfratz:1993sp}, where all lattice artifacts are removed. As a general lesson, our study shows a new way in which ML can enhance particle physics simulations.

\section{Acknowledgements}
\begin{acknowledgments}
K.H.~wishes to thank the AEC and ITP at the University of Bern for their support. U.W.~acknowledges funding from the Swiss National Science Foundation (SNSF) project No.~200020\_208222.
A.I.~and D.M.~acknowledge funding from the Austrian Science Fund (FWF) projects
P 32446, P 34455, P 34764 and PAT3667424.
K.H.~acknowledges funding from the US National Science Foundation under grants 2014150 and 2412320. 
The computational results presented have been achieved in part using the Vienna Scientific Cluster (VSC) and LEONARDO at CINECA, Italy, via an AURELEO (Austrian Users at LEONARDO supercomputer) project, and UBELIX, 
the HPC cluster at the University of Bern.
\end{acknowledgments}

\bibliography{fp_mc} 

\section{Supplemental Material}
\label{sec:supplementaries}

\subsection{Simulation details and gradient-flow measurements}
Here we provide information on the ensembles generated using MC simulations of the FP action. 
The HMC integrator evolves up to trajectory length $\tau=3.0$ using $N_\tau=32$ steps for all ensembles. We use the 4MN4FP variant of Omelyan integration scheme as described in~\cite{Takaishi:2005tz}, generating $N_\mathrm{traj}$ trajectories. The simulation code including the L-CNN for the FP action employed in this work is publicly available \cite{lge-cnn}.

In Table \ref{tab:ensembles_overview_t0.3} we provide for each ensemble the linear lattice size $N_\mathrm{latt}$, the FP action $\beta$-value, and the number of thermalized trajectories  $N_\mathrm{traj}$.  We use symmetric lattice volumes $N_\mathrm{latt}^4$ throughout. In addition, we list the measured gradient-flow observables $t_{0.3}$, $t_{0.4}$, $t_{0.5}$ and $w_{0.3}$ in lattice units.  For the gradient flow we use the third-order Runge-Kutta scheme~\cite{Luscher:2010iy} with timestep $\Delta t_f = 0.01$ on all ensembles. We also list in Table \ref{tab:ensembles_overview_t0.3} the lattice volume $L$ and lattice spacing $a$ in physical units, computed using $\sqrt{t_{0.3}} = 0.1679 \, \mathrm{fm}$~\cite{Giusti:2018cmp}.

\begin{table*}
\caption{Overview of parameters for the FP ensembles used in this work together with a selection of gradient-flow scales in lattice units. The linear lattice extent $L$ and the lattice spacing $a$ are obtained using  $\sqrt{t_{0.3}} = 0.1679 \, \mathrm{fm}$~\cite{Giusti:2018cmp}. 
\label{tab:ensembles_overview_t0.3}}
\begin{ruledtabular}
\begin{tabular}{llrllllll}
$N_\mathrm{latt}$ & $\beta$ & $N_\mathrm{traj}$ & $t_{0.3}/a^2$ & $w_{0.3}/a$ & $t_{0.4}/a^2$ & $t_{0.5}/a^2$ & $L/\mathrm{fm}$ & $a/\mathrm{fm}$\\
10 & $2.65$ & 51232 & $1.2321(27)$ & $1.1455(15)$  & $1.6526(37)$ &  $2.0342(49)$ &$1.5126(16)$ & $0.15126(16)$ \\
10 & $2.75$ & 76128 & $1.5709(35)$ & $1.2848(19)$  & $2.1000(50)$ &  $2.5843(68)$ &$1.3396(15)$ & $0.13396(15)$ \\
16 & $2.75$ & 68256 & $1.5702(13)$ & $1.28442(67)$ & $2.0989(18)$ &  $2.5822(24)$ &$2.1438(9)$  & $0.13399(6)$ \\
14 & $2.80$ & 79392 & $1.7655(19)$ & $1.3606(10)$  & $2.3594(30)$ &  $2.9034(39)$ &$1.7690(10)$ & $0.12636(7)$ \\
14 & $2.85$ & 74400 & $1.9798(26)$ & $1.4394(13)$  & $2.6448(39)$ &  $3.2553(51)$ &$1.6706(11)$ & $0.11933(8)$ \\
14 & $2.90$ & 82368 & $2.2028(31)$ & $1.5171(14)$  & $2.9417(46)$ &  $3.6211(62)$ &$1.5837(11)$ & $0.11312(8)$ \\
16 & $2.90$ & 81600 & $2.2085(25)$ & $1.5192(12)$  & $2.9493(37)$ &  $3.6299(48)$ &$1.8077(10)$ & $0.11298(6)$ \\
14 & $2.95$ & 81312 & $2.4687(44)$ & $1.6067(19)$  & $3.2980(65)$ &  $4.0604(87)$ &$1.4960(14)$ & $0.10686(10)$ \\
16 & $2.95$ & 70336 & $2.4667(33)$ & $1.6062(15)$  & $3.2954(51)$ &  $4.0579(65)$ &$1.7104(11)$ & $0.10690(7)$ \\
16 & $3.00$ & 69696 & $2.7433(43)$ & $1.6923(19)$  & $3.6631(67)$ &  $4.5082(87)$ &$1.6219(13)$ & $0.10137(8)$ \\
16 & $3.05$ & 69600 & $3.0379(48)$ & $1.7809(19)$  & $4.0571(76)$ &  $4.995(10)$  &$1.5413(13)$ & $0.09633(8)$ \\
18 & $3.05$ & 76896 & $3.0435(38)$ & $1.7828(16)$  & $4.0651(59)$ &  $5.0050(81)$ &$1.7323(11)$ & $0.09624(6)$ \\
18 & $3.10$ & 79808 & $3.3840(52)$ & $1.8799(20)$  & $4.5197(80)$ &  $5.565(11)$  &$1.6429(13)$ & $0.09127(7)$ \\
18 & $3.15$ & 66688 & $3.7456(76)$ & $1.9777(28)$  & $5.003(11)$  &  $6.162(16)$  &$1.5615(16)$ & $0.08675(9)$ \\
18 & $3.20$ & 68608 & $4.1508(83)$ & $2.0811(30)$  & $5.542(13)$  &  $6.823(18)$  &$1.4834(14)$ & $0.08241(8)$ \\
\end{tabular}
\end{ruledtabular} 
\end{table*}

In Table \ref{tab:flow_t1_t2} we compile the FP measurements of flow-time ratios $t_{c} / t_{0.3}$ and $t_{c} / w_c^2$ for $c=0.3, 0.4, 0.5$ and the $\beta$-function at renormalized coupling $g_\textrm{GF}^2 = 15.79$ which corresponds to a flow time of $t_{0.3}$.
\begin{table*}
\caption{FP results for gradient-flow observables. The $\beta$-function in the last column is given at the renormalized coupling $g_\textrm{GF}^2 = 15.79$ which corresponds to a
flow time of $t_{0.3}$.\label{tab:flow_t1_t2}}
\begin{ruledtabular}
\begin{tabular}{llllllll}
$N_\mathrm{latt}$ & $\beta$ & $t_{0.3}/w_{0.3}^2 $ & $t_{0.4}/w_{0.4}^2$ & $t_{0.5}/w_{0.5}^2$ & $t_{0.4}/t_{0.3}$ & $t_{0.5}/t_{0.3}$ & $\beta\mathrm{-fn}$ \\
10 & $2.65$ & $0.93909(81)$ & $1.02812(89)$ & $1.07663(97)$ & $1.34137(46)$ & $1.6511(10)$  & $14.401(15)$ \\
10 & $2.75$ & $0.9518(11)$  & $1.0319(12)$  & $1.0769(12)$  & $1.33682(59)$ & $1.6451(12)$  & $14.740(21)$\\
16 & $2.75$ & $0.95177(35)$ & $1.03223(38)$ & $1.07802(43)$ & $1.33672(19)$ & $1.64455(43)$ & $14.7399(69)$ \\
14 & $2.80$ & $0.95372(56)$ & $1.03181(61)$ & $1.07686(66)$ & $1.33635(30)$ & $1.64453(66)$ & $14.798(11)$ \\
14 & $2.85$ & $0.95556(66)$ & $1.03158(68)$ & $1.07618(72)$ & $1.33592(34)$ & $1.64428(75)$ & $14.847(13)$ \\
14 & $2.90$ & $0.95706(70)$ & $1.03175(75)$ & $1.07602(76)$ & $1.33550(38)$ & $1.64385(79)$ & $14.885(14)$ \\
16 & $2.90$ & $0.95696(54)$ & $1.03216(59)$ & $1.07645(60)$ & $1.33542(30)$ & $1.64354(64)$ & $14.882(11)$ \\
14 & $2.95$ & $0.95640(79)$ & $1.03112(84)$ & $1.07544(89)$ & $1.33586(41)$ & $1.64458(91)$ & $14.875(15)$ \\
16 & $2.95$ & $0.95615(65)$ & $1.03066(69)$ & $1.07436(74)$ & $1.33601(34)$ & $1.64522(77)$ & $14.870(13)$ \\
16 & $3.00$ & $0.95784(76)$ & $1.03229(83)$ & $1.07644(82)$ & $1.33520(41)$ & $1.64330(88)$ & $14.907(15)$ \\
16 & $3.05$ & $0.95780(82)$ & $1.03129(86)$ & $1.07502(95)$ & $1.33550(43)$ & $1.64425(94)$ & $14.911(16)$ \\
18 & $3.05$ & $0.95749(67)$ & $1.03108(69)$ & $1.07495(73)$ & $1.33562(34)$ & $1.64446(75)$ & $14.904(13)$ \\
18 & $3.10$ & $0.95747(75)$ & $1.03109(80)$ & $1.07496(81)$ & $1.33559(40)$ & $1.64445(88)$ & $14.904(15)$ \\
18 & $3.15$ & $0.95750(92)$ & $1.0303(10)$  & $1.0739(10)$  & $1.33581(49)$ & $1.6451(11)$  & $14.909(18)$ \\
18 & $3.20$ & $0.95841(99)$ & $1.0316(11)$  & $1.0755(12)$  & $1.33519(53)$ & $1.6437(12)$  & $14.927(19)$ \\
\end{tabular}
\end{ruledtabular}
\end{table*}

Autocorrelation times $\tau_c$ of the action density for $c=0.3, 0.4, 0.5$ are estimated through single exponential fits of the autocorrelation function of $t^2_c E(t_c)$, combining $N_\mathrm{chains}$ independent runs of unequal length, the shortest of which has length $N^*_\mathrm{s}$, cf.~Table~\ref{tab:auto}.\footnote{We note that the different number of chains $N_\mathrm{chains}$ and minimal chain lengths $N^*_\mathrm{s}$ are simply due to practical constraints, such as available computing time on various GPU architectures, and have no relevance for the autocorrelation times quoted in Table~\ref{tab:auto} or the results in Tables \ref{tab:ensembles_overview_t0.3} and \ref{tab:flow_t1_t2}.}

Gradient-flow measurements are made every $N_\mathrm{skip}=32$ trajectories, except for the ensemble marked $\dagger$ with more frequent measurements at $N_\mathrm{skip}=4$. The measurements in ensembles at $\beta=2.80, 2.75, 2.65$ show no significant autocorrelations, which is why the corresponding entries are missing from the Table. Finally, the autocorrelation times, as tabulated in Table~\ref{tab:auto}, are reexpressed in units of HMC trajectories.

The autocorrelation times are in general short, growing somewhat towards finer lattice spacing, but are significantly less than 32 trajectories. In the statistical analysis, gradient-flow measurements are further blocked with $N_\mathrm{block}=8$ to suppress any remnant autocorrelation effects in the data, and we use $N_\mathrm{resample}=2000$ in the bootstrap determination of statistical errors. 
\begin{table}
\caption{Statistics and autocorrelation times $\tau_c$ of $t_c^2 E(t_c)$ in units of HMC trajectories for the FP ensembles with $a < 0.12$ fm. The ensembles are assembled from $N_\mathrm{chains}$ independent chains of unequal length, the shortest of which has length $N^*_\mathrm{s}$. Gradient-flow measurements are done every $N_\mathrm{skip}=32$ trajectories, except for the ensemble marked $\dagger$ with $N_\mathrm{skip}=4$. Autocorrelation times for ensembles with $\beta \leq 2.800$ are too short to be measured accurately.
\label{tab:auto}}
\begin{ruledtabular}
\begin{tabular}{lrrrrrr}
$\beta$ & $N_\mathrm{latt}$ & $N_\mathrm{chains}$ & $N^*_\mathrm{s}$ & $\tau_{0.3}$ & $\tau_{0.4}$ & $\tau_{0.5}$ \\
2.650 & 10 & 2 & 25536 & - & - & - \\
2.750 & 10 & 3 & 25280 & - & - & - \\
2.750 & 16 & 24 & 2656 & - & - & - \\
2.800 & 14 & 8 & 9408 & - & - & - \\
2.850$\dagger$ & 14 & 1 & 10844 & 7.1 & 7.6 & 7.8 \\
2.850 & 14 & 8 & 2496 & 6.9 & 6.4 & 6.2 \\
2.900 & 14 & 8 & 10176 & 9.3 & 9.7 & 9.9 \\
2.900 & 16 & 16 & 3776 & 8.8 & 9.3 & 9.4 \\
2.950 & 14 & 8 & 9888 & 12.8 & 13.3 & 13.4 \\
2.950 & 16 & 16 & 2496 & 9.4 & 9.9 & 10.2 \\
3.000 & 16 & 16 & 3520 & 14.7 & 15.2 & 15.5 \\
3.050 & 16 & 16 & 3168 & 11.3 & 12.0 & 12.5 \\
3.050 & 18 & 28 & 2688 & 14.2 & 15.1 & 15.6 \\
3.100 & 18 & 29 & 2720 & 16.4 & 17.0 & 17.6 \\
3.150 & 18 & 32 & 1632 & 19.8 & 20.7 & 21.3 \\
3.200 & 18 & 32 & 2112 & 17.3 & 19.2 & 20.7 \\
\end{tabular}
\end{ruledtabular}
\end{table}

\clearpage

\begin{figure}[!htbp]
    \centering
    \includegraphics[width=.95\linewidth]{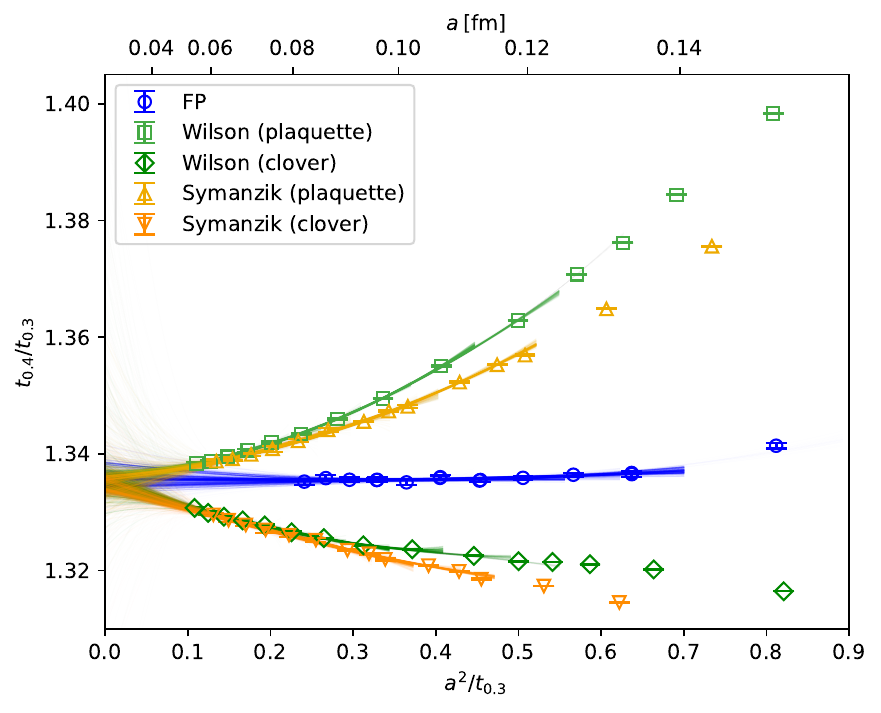}
    \includegraphics[width=.95\linewidth]{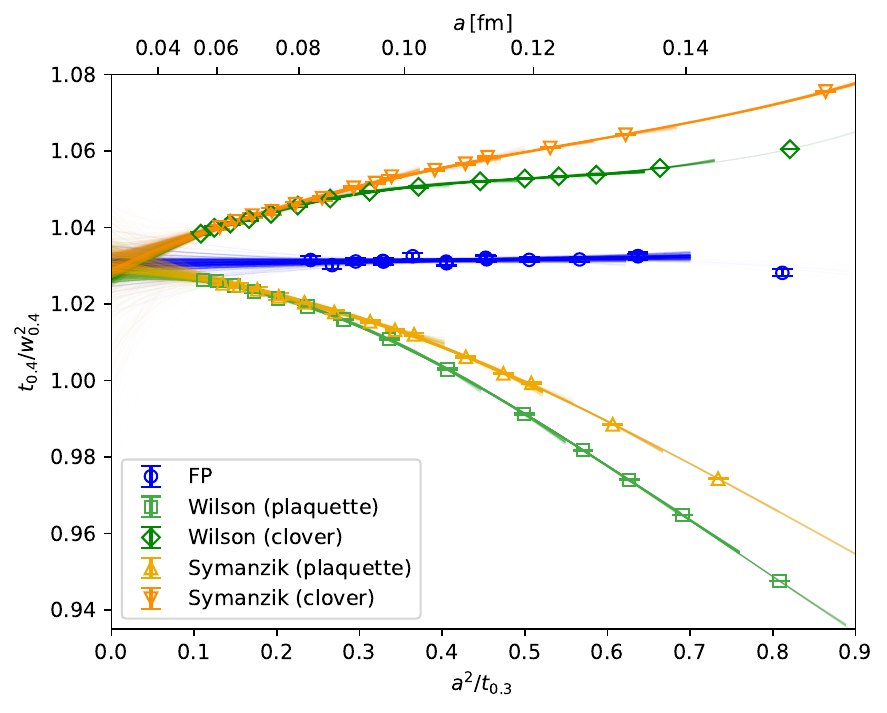}
    \includegraphics[width=.95\linewidth]{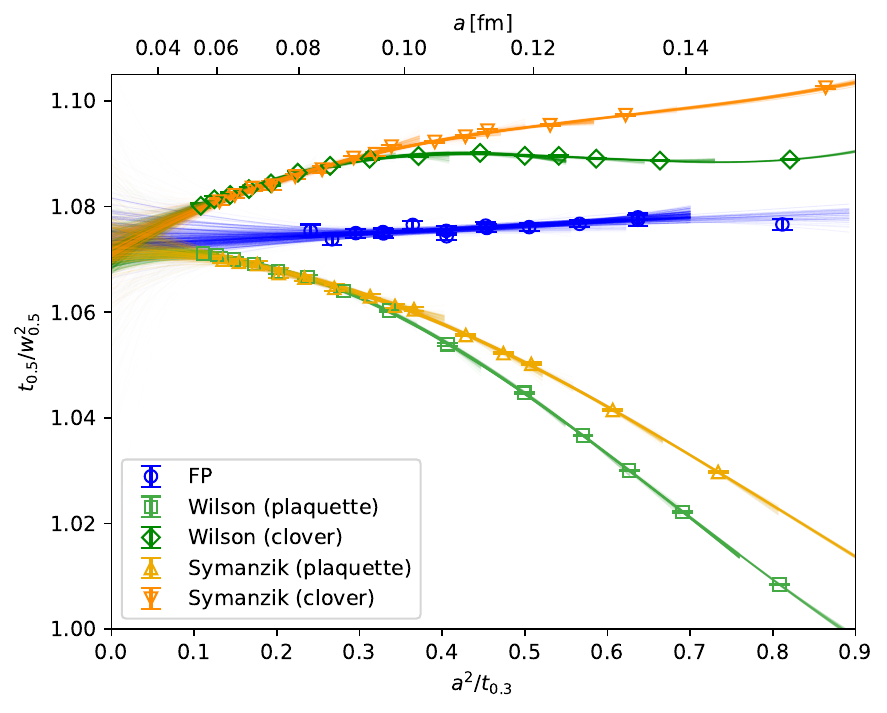}
    \caption{Continuum-limit extrapolations of FP, Wilson and Symanzik action measurements, for the ratios $t_{0.4}/t_{0.3}$ ({\it top}), $t_{0.4}/w_{0.4}^2$ ({\it middle}), and $t_{0.5}/w_{0.5}^2$ ({\it bottom}). Lattice artifacts are small with the FP action and gradient flow (below 1\% at $a \simeq 0.14$ fm). A variety of polynomial fits with different ranges and highest power $(a^2/t_0)^n$ are shown, shaded according to their respective AIC weight.}
    \label{fig:FPmore}
\end{figure}

\subsection{Continuum limits and AIC averaging procedure}

We perform the continuum limit for each dimensionless quantity $z$
by fitting the results from the simulations with polynomials of order $n$ in $a^2/t_{0.3}$. For the FP results we use $n=0, 1$, and $2$, while for the results from the Wilson and tree-level Symanzik improved action we use $n=1,2$, and $3$. The results obtained with the pla\-quet\-te and clover discretization of the action density $E$ are treated as independent and we do not perform combined continuum limits. 

\begin{figure*}[!htbp]
\begin{minipage}[t]{0.48\textwidth}
    \centering
    \includegraphics[width=.95\linewidth]{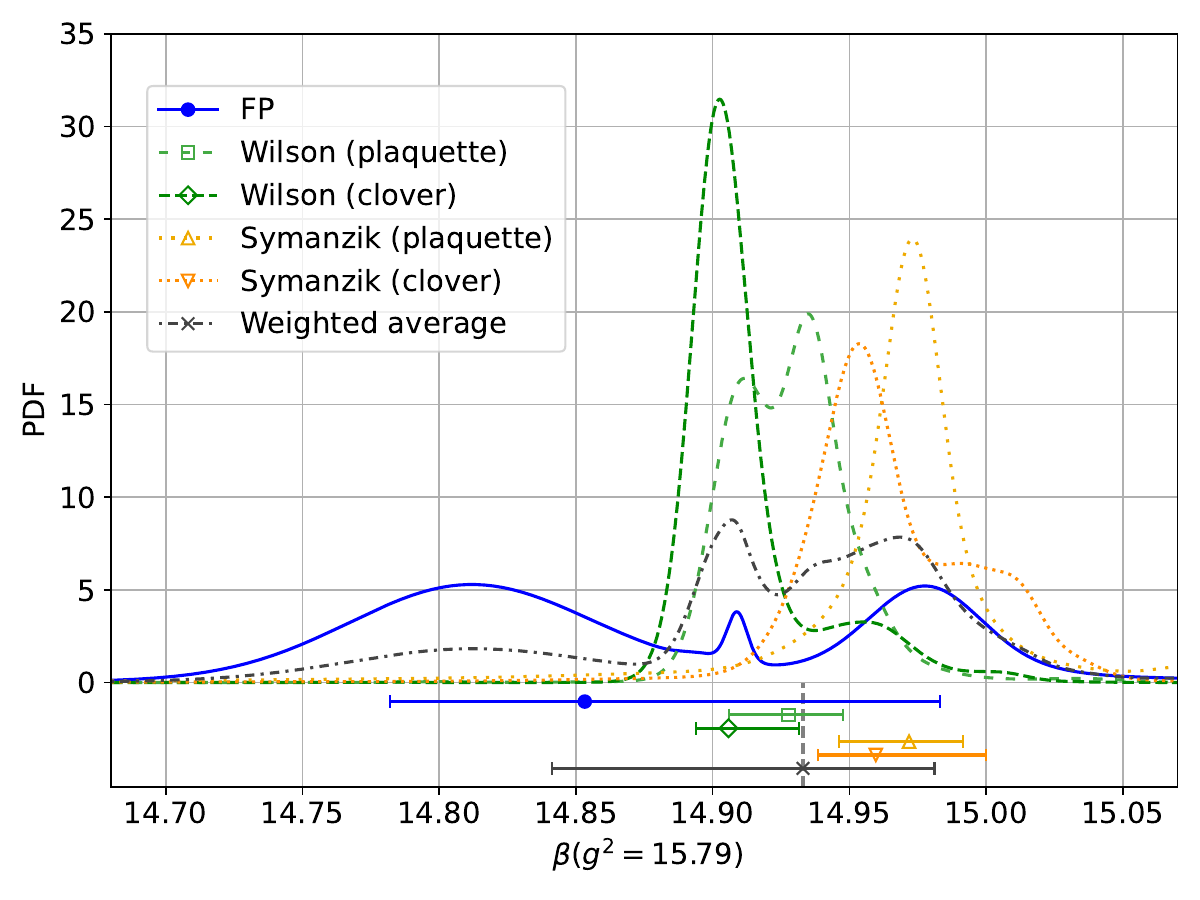}
    \includegraphics[width=.95\linewidth]{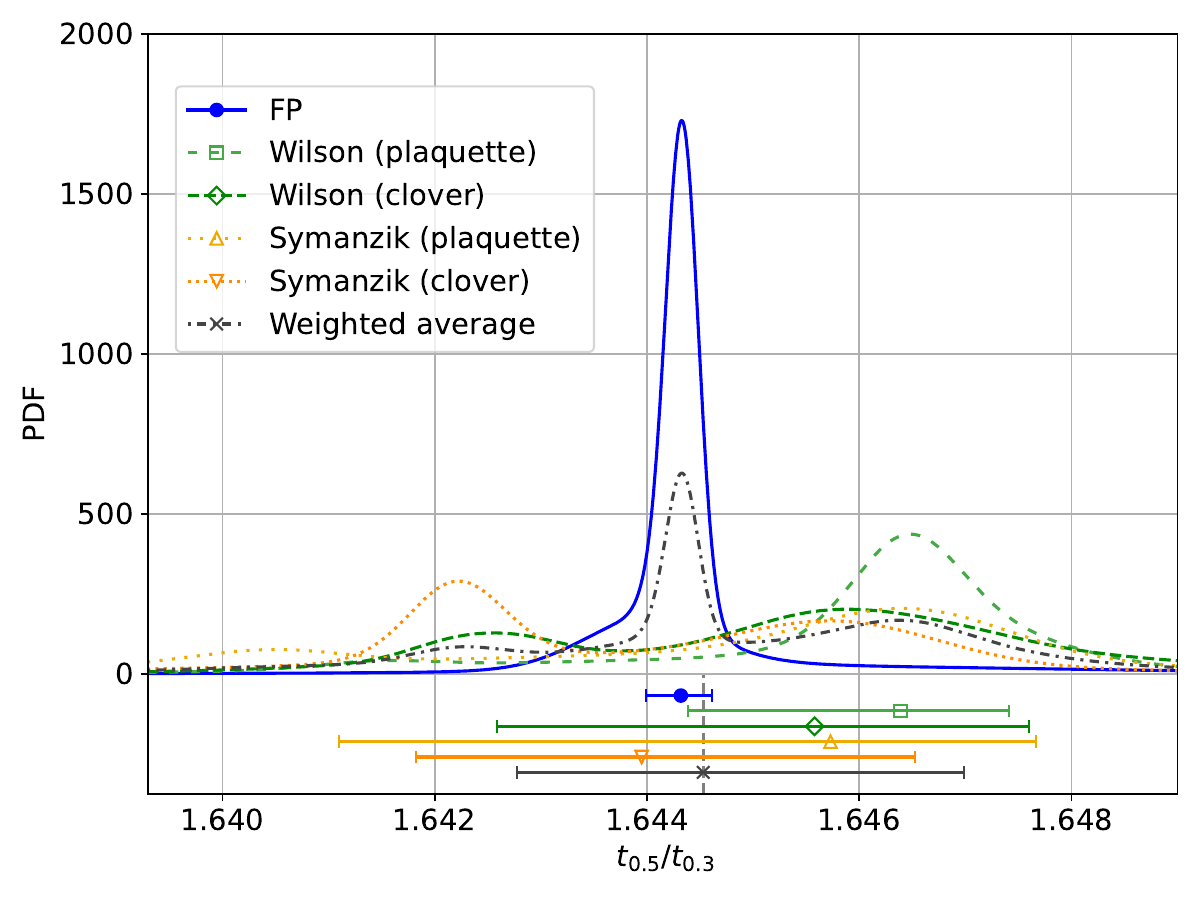}
    \includegraphics[width=.95\linewidth]{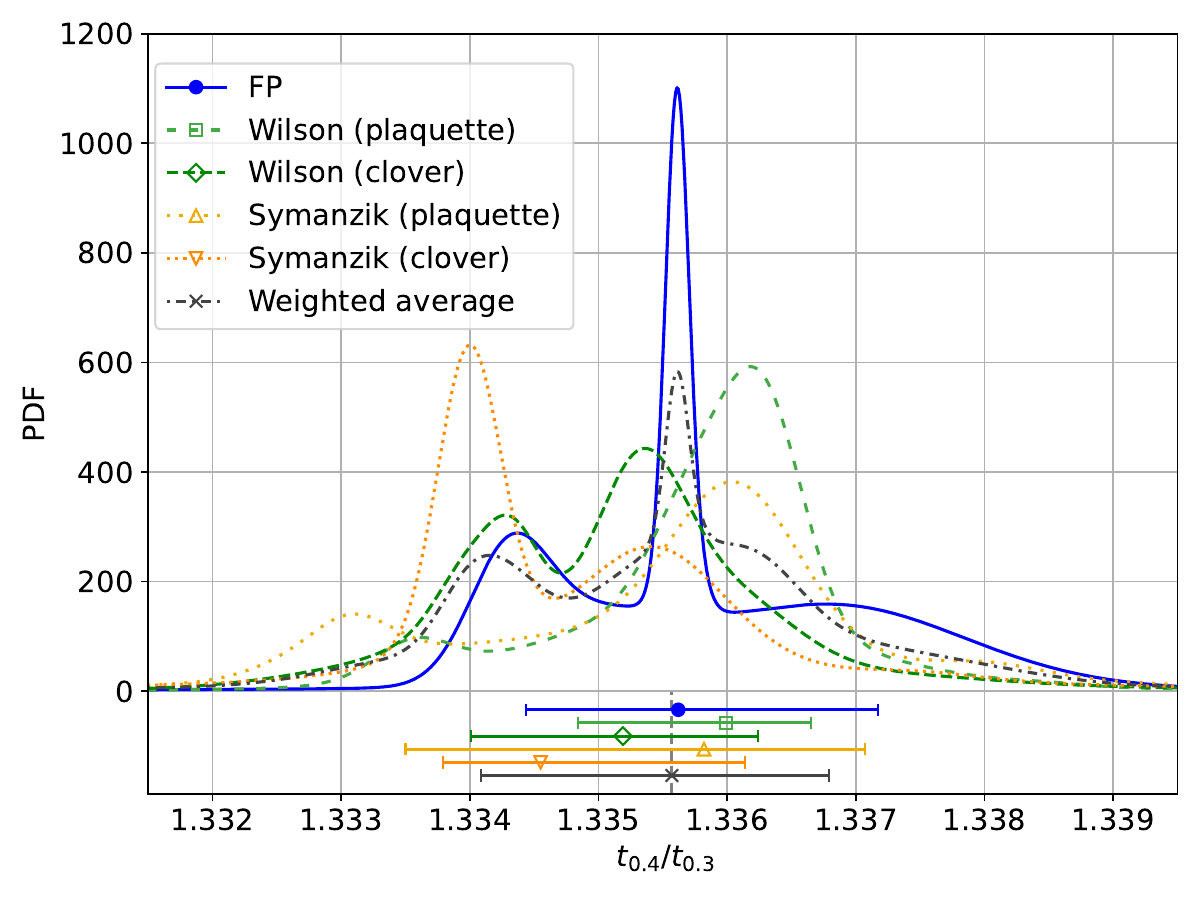}
    \caption{Comparison of various PDFs for the quantity $\beta(g^2=15.79$ ({\it top}), $t_{0.5}/t_{0.3}$ ({\it middle}), $t_{0.4}/t_{0.3}$ ({\it bottom}).  Below the plot, the median, 16$^{\rm th}$ and 84$^{\rm th}$ percentiles are marked.}
    \label{fig:PDF_comparison_1}
\end{minipage}
\hfill
\begin{minipage}[t]{0.48\textwidth}
    \centering
    \includegraphics[width=.95\linewidth]{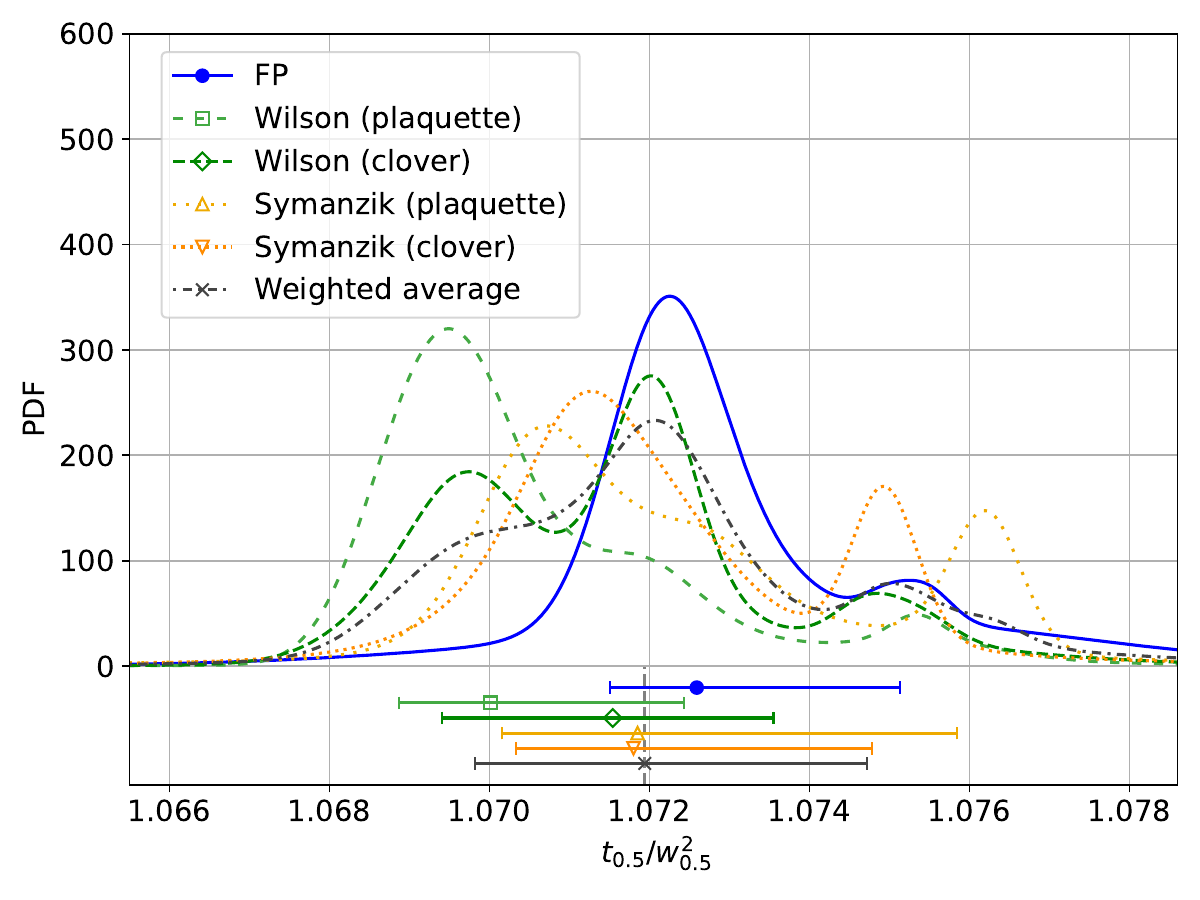}
    \includegraphics[width=.95\linewidth]{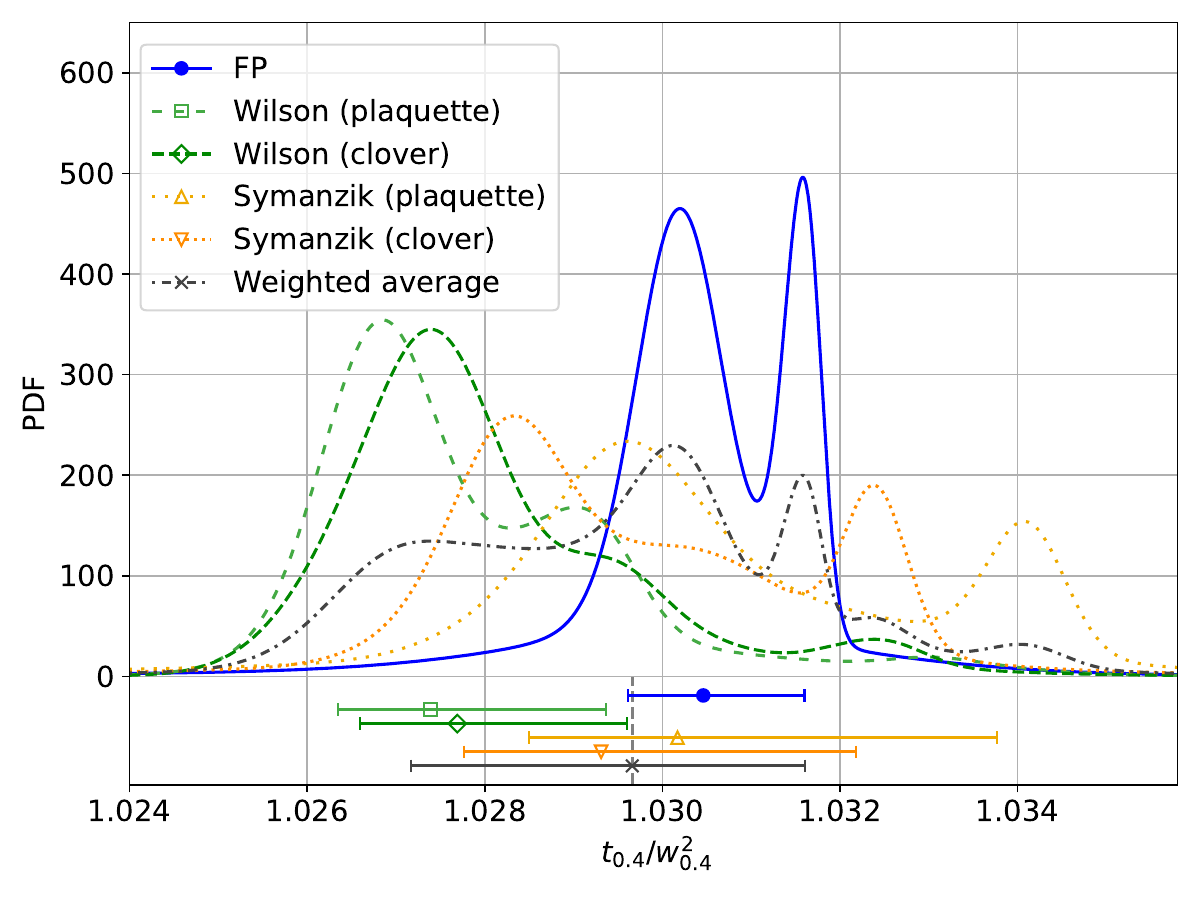}
    \includegraphics[width=.95\linewidth]{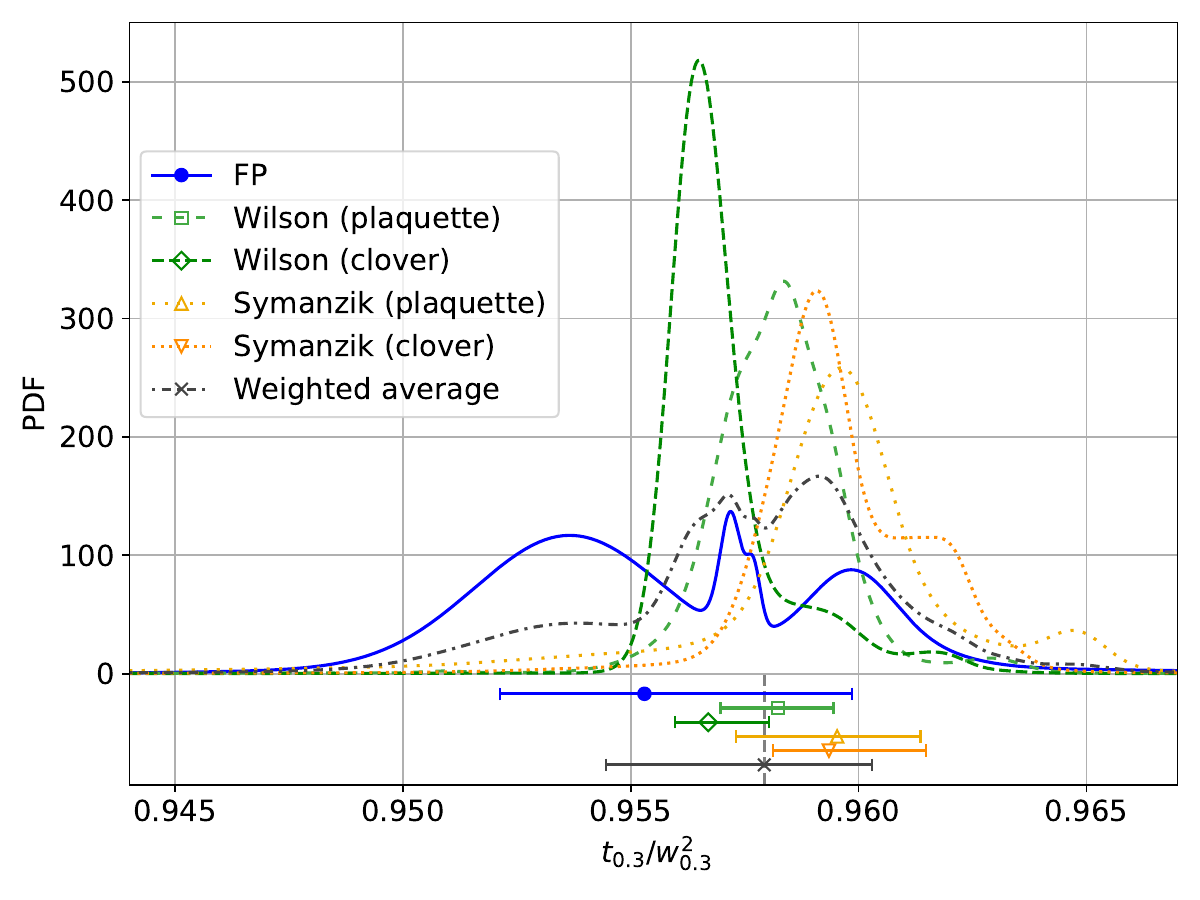}
    \caption{Comparison of various PDFs for the quantity $t_{0.5}/w_{0.5}^2$ ({\it top}), $t_{0.4}/w_{0.4}^2$ ({\it middle}), $t_{0.3}/w_{0.3}^2$ ({\it bottom}).  Below the plot, the median, 16$^{\rm th}$ and 84$^{\rm th}$ percentiles are marked.}
    \label{fig:PDF_comparison_2}
\end{minipage}
\end{figure*}

The continuum limits are illustrated in Figures \ref{fig:t0w0ratio}, \ref{fig:betafn}, and \ref{fig:FPmore} where we show the results of the simulations together with a selection of (bootstrapped) polynomial
fits indicating the included data points by the extent of the lines and shaded according to their Akaike Information Criterion (AIC) weight~\cite{Akaike:1974vps} calculated as follows. 
For each possible fit $i$ given $n_{\rm tot}$ data of which $n$ are fitted using $k$ parameters, the  weight $\omega_i$ for the continuum limit $z_i$ is given by
\begin{align}
    {\rm AIC_i} &= \chi_i^2 + 2k + 2(n_{\rm tot}-n),\\
    \omega_i &= \exp(-{\rm AIC_i}/2) .
\end{align}
The AIC weights are also used to estimate systematic effects in the continuum extrapolation as follows. The probability distribution function (PDF) of $z$ is built from a weighted combination of each distribution of $z_i$ parameterized as a normal distribution ${\cal N}(z_i,\sigma_i)$. The median of the resulting PDF is taken as the final result, and the 16$^{\rm th}$ and 84$^{\rm th}$ percentiles of the cumulative distribution function as (possibly asymmetric) uncertainty ranges. This procedure combines both statistical and systematic errors in the extrapolation. In Figures \ref{fig:PDF_comparison_1} and \ref{fig:PDF_comparison_2} we show the PDFs for each continuum limit and quantity. 
The multipeak structures seen in the PDFs reflect the fact that at the given high statistical accuracy the continuum extrapolations in general depend on the order of the fitted polynomial. As a consequence, the total error is dominated by the systematics of the continuum extrapolation as pointed out in the introduction. The final results from the continuum extrapolations are collected in Table \ref{tab:flow_extrapolation_UW_KH2_new}.

In comparing the results, 
we calculate the weighted average $\mathrm{PDF}_{\mathrm{avg}}$ from the individual PDFs using
\begin{equation}
    \mathrm{PDF}_{\mathrm{avg}} = \frac{\sum_k \omega_k \mathrm{PDF}_k}{\sum_k \omega_k}, 
\end{equation}
where the index $k$ runs over the five cases: FP, Wilson (plaquette), Wilson (clover), Symanzik (plaquette), and Symanzik (clover). $\mathrm{PDF}_k$ denotes the respective individual PDF. Since the data from the plaquette and the clover definition of the action density for either the Wilson or the Symanzik action are highly correlated, we follow a conservative approach and suppress these contributions by a factor of $\omega_k = 1/2$, while $\omega_\mathrm{FP}=1$. Finally, Figure~\ref{fig:FPwilson} shows the deviation $z_k - z_{\mathrm{avg}}$, where $z_{\mathrm{avg}}$ is the median of 
$\mathrm{PDF}_{\mathrm{avg}}$, for all five cases, together with their respective errors.

\begin{table*}
\caption{
Continuum values of gradient-flow observables using AIC model averaging with fits up to $a^6$-terms (Wilson and Symanzik) and up to $a^4$-terms (FP).
\label{tab:flow_extrapolation_UW_KH2_new}}
\begin{ruledtabular}
\begin{tabular}{lllllll}
Observable & FP & Wilson clover & Wilson plaquette & Symanzik clover & Symanzik plaquette & Weighted average\\
$t_{0.3}/w_{0.3}^2$ & $0.9553(^{+46}_{-32})$ & $0.9567(^{+13}_{\phantom{1}-7})$ & $0.9582(^{+12}_{-13})$ & $0.9594(^{+21}_{-12})$ & $0.9595(^{+18}_{-22})$ & $0.9579(^{+24}_{-35})$  \\
$t_{0.4}/w_{0.4}^2$ & $1.0305(^{+11}_{\phantom{0}-9})$ & $1.0277(^{+19}_{-11})$ & $1.0274(^{+20}_{-10})$ & $1.0293(^{+29}_{-15})$ & $1.0302(^{+36}_{-17})$ & $1.0297(^{+19}_{-25})$  \\
$t_{0.5}/w_{0.5}^2$ & $1.0726(^{+25}_{-11})$ & $1.0715(^{+20}_{-21})$ & $1.0700(^{+24}_{-11})$ & $1.0718(^{+30}_{-15})$ & $1.0718(^{+40}_{-17})$ & $1.0719(^{+28}_{-21})$  \\
$t_{0.4}/t_{0.3}$ & $1.3356(^{+15}_{-12})$ & $1.3352(^{+10}_{-12})$ & $1.3360(^{\phantom{0}+7}_{-11})$ & $1.3345(^{+16}_{\phantom{0}-8})$ & $1.3358(^{+12}_{-23})$ & $1.3356(^{+12}_{-15})$  \\
$t_{0.5}/t_{0.3}$ & $1.6443(^{+3}_{-3})$ & $1.6456(^{+20}_{-30})$ & $1.6464(^{+10}_{-20})$ & $1.6440(^{+26}_{-21})$ & $1.6457(^{+19}_{-46})$ & $1.6445(^{+25}_{-18})$  \\
$\beta(g^2 = 15.79)$ & $14.853(^{+130}_{\phantom{0}-71})$ & $14.906(^{+26}_{-12})$ & $14.928(^{+20}_{-22})$ & $14.960(^{+40}_{-21})$ & $14.972(^{+20}_{-26})$ & $14.933(^{+48}_{-92})$  \\
\end{tabular}
\end{ruledtabular}
\end{table*}

\end{document}